\documentclass[twocolumn,english,prl,floatfix,showpacs]{revtex4-1}

\usepackage[T1]{fontenc}
\usepackage[latin9]{inputenc}
\usepackage{amsmath}
\usepackage{amssymb}
\usepackage{graphicx}
\usepackage{wasysym}
\usepackage{caption}
\usepackage{subcaption}

\makeatletter

\graphicspath{ {./figures/} }

\usepackage{babel}

\@ifundefined{showcaptionsetup}{}{%
 \PassOptionsToPackage{caption=false}{subfig}}
\usepackage{subfig}
\makeatother

\usepackage{babel}
\begin{document}

\title{The Griffiths Phase on Hierarchical Modular Networks with Small-world
Edges}

\author{Shanshan Li}

\affiliation{Department of Physics, Emory University, Atlanta, GA 30322; USA }
\begin{abstract}
The Griffiths phase has been proposed to induce a stretched critical
regime that facilitates self-organizing of brain networks for optimal
function. This phase stems from the intrinsic structural heterogeneity
of brain networks, such as the hierarchical modular structure. In
this work, we extend this concept to modified hierarchical networks
with small-world connections based on Hanoi networks. 
Through extensive simulations, we identify the role of
an exponential distribution of the inter-moduli connectivity probability
across hierarchies determining the emergence of the Griffiths phase.
Numerical results and the complementary spectral analysis on the relevant networks 
can be helpful for a deeper understanding of the essential structural characteristics of
finite dimensional networks to support the Griffiths phase.  
\end{abstract}


\maketitle

\section{Introduction\label{sec:Introduction}}

The Griffiths phase (GP) is characterized by generic power-laws over
a broad region in the parameter space. It provides an alternative
mechanism for critical behavior in brain networks without fine tuning
\citep{moretti2013griffithsonHMNs,odor2015griffithsinHMNs}. The well-known
criticality hypothesis suggests biological systems operate at the
borderline between the sustained active and inactive state. It has
been observed in various processes such as gene expression \citep{nykter2008gene},
cell growth \citep{kaneko2012cellgrowth} and neuronal avalanches
\citep{beggs2003avalanches}. The critical point enables optimal transmission
and storage of information \citep{plenz2007optimal,beggs2008optimal},
maximal sensitivity to stimuli \citep{kinouchi2006maximal}, optimal
computational capabilities \citep{legenstein2007computational}. Empirical
studies on brain networks \citep{barbieri2012broadcriticality,rubinov2011broadcriticality,wang2012broadcriticality},
however, exhibit a broad critical region. It is confirmed numerically
and analytically that the structural heterogeneity induces the Griffiths
phase that eventually enhances the self-organization mechanism of
brain networks.

Brain networks have been found to be organized into moduli across
hierarchies \citep{sporns2004BrainNetworks,meunier2010BrainNetworks,kaiser2011BrainNetworks}.
Moduli in each hierarchy are grouped into larger moduli, forming
a fractal-like structure. Previous work models brain networks with
finite dimensional hierarchical modular networks (HMNs) \citep{moretti2013griffithsonHMNs,odor2015griffithsinHMNs},
and successfully confirms the existence of the Griffiths phase using
dynamical models, such as the Susceptible-Infected-Susceptible (SIS)
model and the Contact Process (CP). The essential characteristics
of previous network models is an exponential distribution of inter-moduli
connectivity probability across hierarchies that eventually leads
to an exponential distribution of moduli size. It is conjectured that plain modular networks
are not able to support the Griffiths phase, disorder in different scales significantly influences properties
of critical behaviors \citep{moretti2013griffithsonHMNs}. In this work, we extend the idea of a Griffith
phase to other hierarchical structures encountered in previous studies on dynamical processes on complex networks. 

Certain hierarchical networks, with a self-similar structure and small-world connections, have shown to exhibit novel dynamics \citep{boettcher2009patchy,berker2009critical,boettcher2011fixed,boettcher2012ordinary,singh2014scaling,singh2014explosive}. Here, we design hierarchical models based on one such example, the Hanoi networks  \citep{boettcher2008hierarchical, boettcher2009patchy, boettcher2011fixed, boettcher2015real}.
To tune the modular feature that is present in brain networks, we modify a single node of the original network into a fully connected clique with a varying size. By introducing different kinds of inter-moduli connections, we explore the essential heterogeneous connectivity pattern to induce the Griffiths phase on finite dimensional networks. We find that an exponential distribution of the inter-moduli connectivity probability
across hierarchies plays an essential role affecting the property of the phase transition at criticality. As a complement to the computational approach, the spectral analysis on the adjacency matrix of networks is conducted. A localized principle eigenvector of the network adjacency matrix indicates the network heterogeneity, which has been used to quantify the localization of activity on networks \citep{goltsev2012localization}. This concept has been applied to analytically explain the emergence of rare regions and the Griffiths phase \citep{moretti2013griffithsonHMNs,odor2015griffithsinHMNs,odor2013spectral}. The observation that a localized principle eigenvector is not necessarily the fingerprint of the Griffiths phase has been found in highly-connected networks with intrinsic weight disorder or finite-size random networks with power-law degree distributions \citep{odor2015griffithsinHMNs,odor2016finitesize}. As an extension to finite dimensional models, we find a class of networks where the Griffiths phase is absent although their principle eigenvectors are localized. 

This paper is organized as follows: in Sec. \ref{sec:Graph-Structure},
we describe the structural properties of hierarchical modular networks
on which we study the SIS model and its critical behavior; in Sec.\ref{sec:SIS-model},
we review the SIS model and the spectral analysis on the network adjacency
matrix, and apply the analytical tool to all the networks we propose;
in Sec.\ref{sec:results}, we present the numerical results for the
SIS model evolving on the networks we consider. We conclude in Sec.\ref{sec:conclusion}
by highlighting the significance of the exponential distribution of
the moduli size or equivalently the inter-moduli connectivity probability
on the emergence of the Griffiths phase.
\section{Network Structure\label{sec:Graph-Structure}}

\begin{figure}
\begin{centering}
\includegraphics[scale=0.3,angle=0]{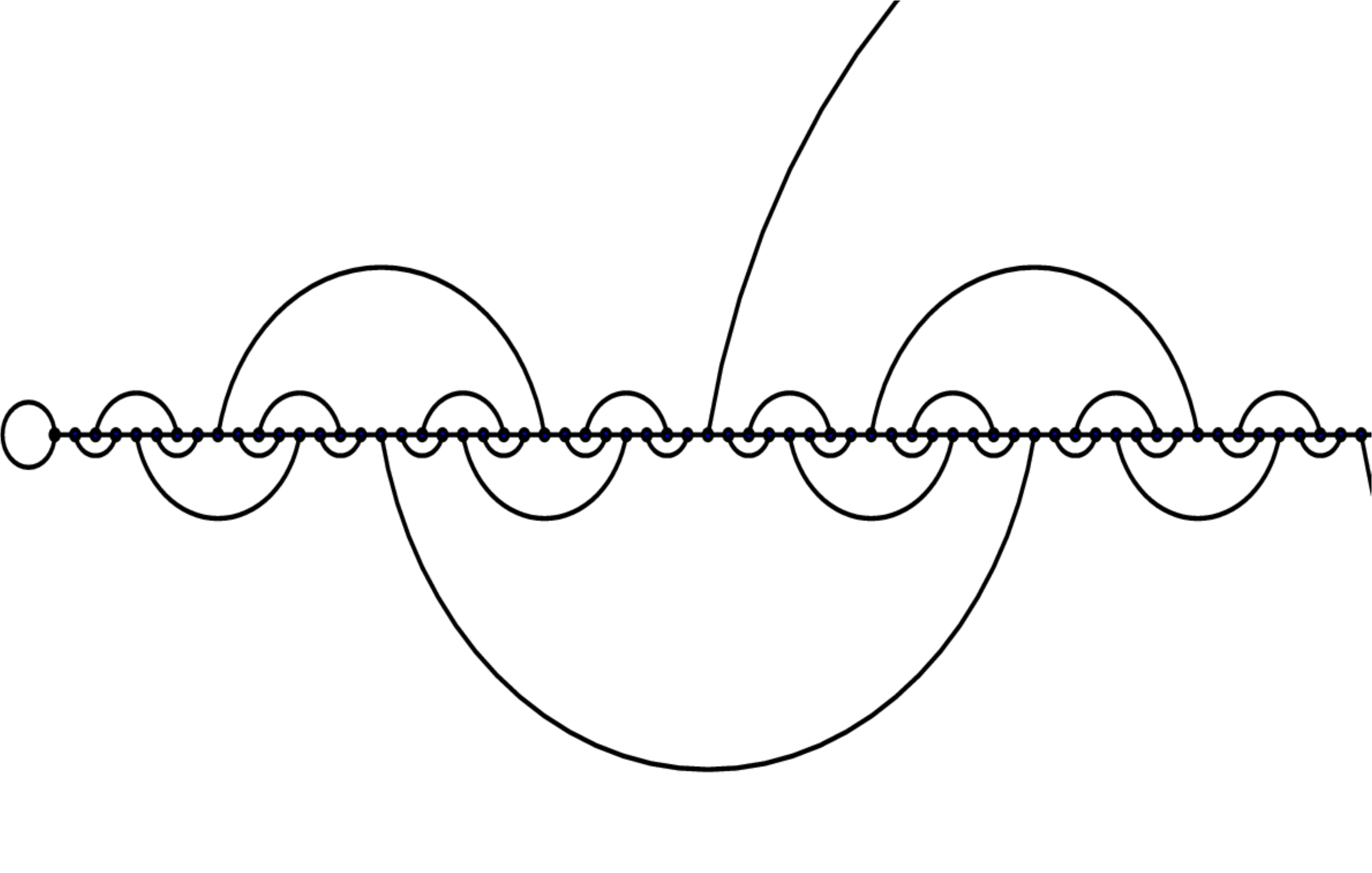}
\end{centering}
\captionsetup{justification=raggedright, singlelinecheck=false}
\caption{\label{fig:hanoi3}Depiction of the Hanoi network of generation $g=6$. The network features a regular geometric structure, in the form of a one dimensional backbone, and a distinct set of recursive small-world links. The node degree is uniformly $3$.}
\end{figure}

The Hanoi networks \citep{boettcher2008hierarchical, boettcher2009patchy, boettcher2011fixed, boettcher2015real}
are based on a simple geometric backbone, a one-dimensional line of
$n=2^{g}$ nodes. Each node is at least connected to its nearest neighbor
left and right on the backbone. To construct the hierarchy to $g$-th
generation, consider parameterizing any node $x<n$ (except for zero)
\emph{uniquely} in terms of two integers $(i,j)$, $i\geq1$ and $1\leq j\leq2^{g-i}$,
via 
\begin{eqnarray}
x & = & 2^{i-1}\left(2j-1\right).\label{eq:numbering}
\end{eqnarray}
Here, $i$ denotes the level of hierarchy whereas $j$ labels consecutive
nodes within each hierarchy. Such a parametrization raises a natural
pattern for long-range small-world edges that are formed by the neighbors
$x=2^{i-1}(4j-3)$ and $y=2^{i-1}(4j-1)$ for $1\leq j\leq2^{g-i-1}$,
as shown in Fig.(\ref{fig:hanoi3}). Eventually, this procedure constructs
a finite dimensional hierarchical network with a uniform finite node degree $3$, and
a diameter of $\sim\sqrt{n}$, which is denoted as HN3 \citep{boettcher2008hierarchical,boettcher2009patchy, boettcher2011fixed}.

To model the modular property of real-world brain networks, we replace
each single node $x$ in HN3 by a fully-connected clique that contains
a finite number $m$ of nodes, thus forming a network with size $n\times m$. Maintaining the structural properties of HN3, the self-similar
structure, and the small-world connections, we design two connectivity
patterns between moduli in the same hierarchy. In the first paradigm,
the single edge in the original HN3 is now formed by two randomly
chosen inter-clique nodes, which we denote as HMN1. The second paradigm is inspired by previous hierarchical modular models \citep{moretti2013griffithsonHMNs,odor2015griffithsinHMNs}.
To distinguish it from HMN1, we denote it as HMN2. Previous models
share a common feature, the exponential distribution of the inter-moduli
connectivity probability. Moduli are connected in either a stochastic
way with a level-dependent probability $p_{i}$ or a deterministic
way with a level-dependent number of edges. 

Since an infinite dimensional network is predicted not to support the Griffiths phase \citep{moretti2013griffithsonHMNs},
to maintain a finite fractal dimension, the size of moduli exponentially
increases as the inter-moduli connectivity probability exponentially decreases.
Here, we use the stochastic scheme to construct HMN2. In HMN2, for
the second hierarchy, the clique $2(2j-1)$ is grouped with the neighbor
clique $2(2j-1)-1$ and $2(2j-1)+1$ forming a moduli. For the third
hierarchy, the clique $2^{2}(2j-1)$ is grouped with three left neighbor
cliques up to the clique $2^{2}(2j-1)-3$ and three right neighbor
cliques up to the clique $2^{2}(2j-1)+3$. Repeating this procedure
to $g$-th generation, the size of moduli of $i$-th generation is
$m(2^{i}-1)$. The number of all possible stochastic connections between
two moduli is $m^{2}\left(4^{i}-2^{i+1}+1\right)$. Thus, to ensure
at least one edge between them, the level-dependent probability $p_{i}$
is at least $1/\left(m^{2}\left(4^{i}-2^{i+1}+1\right)\right)$.

\section{Susceptible-Infected-Susceptible Model and the Spectral Analysis\label{sec:SIS-model}}

Certain fundamental dynamical models, the Susceptible-Infected-Susceptible
(SIS) model and the Contact Process (CP), have been used to model
the activity propagation on brain networks \citep{moretti2013griffithsonHMNs,odor2015griffithsinHMNs}.
Previous studies focus on the emergence of the Griffiths phase on
general complex networks using these simplified models. Quenched disorder,
either intrinsic to nodes or topological, has been shown to smear
the phase transition at critical points and generate the Griffiths
phase. Special $\mathcal{\mathsf{\mathit{rare}}}$ $\mathit{regions}$
(RRs) emerge in this dynamical process evolving on networks with quenched
disorder. Statistically, the active state lingers in these rare regions
for a typical time that grows exponentially with their sizes, and
eventually ends up in the absorbing state \citep{griffiths1969GP,noest1986DisorderPercolation,vojta2006rareregion}.
The emerging exponentially distributed rare regions induce power-law
decays with continuously varying exponents, i.e. the Griffiths phase.

The essential disorder can stem from a node-dependent propagation
rate (intrinsic quenched disorder) \citep{munoz2010griffithsonCNs,juhasz2012rare}.
Recent results also present evidence that the Griffiths phase emerges
due to the quenched disorder on the edges, such as in tree networks
with a correlated weight pattern \citep{odor2012slow} and in random
networks with exponentially suppressed weight scheme \citep{odor2013spectral}.
The Quenched Mean-Field (QMF) approximation applies a spectral analysis
on the network adjacency matrix that analytically explains emerging
rare regions and the Griffiths phase on networks with the quenched
disorder \citep{goltsev2012localization, odor2013spectral}. In absence
of the quenched disorder, the Griffiths phase can also be a consequence
of the structural heterogeneity of finite dimensional networks that
is expected to have a similar role as the quenched disorder \citep{munoz2010griffithsonCNs,moretti2013griffithsonHMNs}.
This analytical procedure successfully confirms the Griffiths phase
on finite dimensional hierarchical modular networks in previous work
\citep{moretti2013griffithsonHMNs}. In this
section, we will focus on the SIS model and apply the spectral analysis
on all the finite dimensional structures we consider.

\subsection{SIS Model and the Simulation\label{subsec:SIS-simulation}}

In SIS model, each node in networks is described by a binary state,
active ($\sigma=1$) or inactive ($\sigma=0$). An active node is
deactivated with a unit rate, while it propagates the activity to
its neighbors with a rate $\lambda$. The evolution equation for the
probability $\rho_{x}\left(t\right)$ that node $x$ is active at
time $t$ is 
\begin{eqnarray}
\frac{d}{dt}\rho_{x}\left(t\right) & = & -\rho_{x}\left(t\right)+\lambda\left[1-\rho_{x}\left(t\right)\right]\sum_{y=1}^{N}A_{xy}\rho_{y}\left(t\right),\label{eq:SIS_eq}
\end{eqnarray}
in which $A_{xy}$ is the network adjacency matrix.

We here briefly introduce the method we use to perform the simulation
for the SIS model. The large-scale numerical simulation method of
the SIS model developed in \citep{ferreira2012epidemic} determines
the critical propagation rate $\lambda_{c}$ efficiently for various
networks. This algorithm considers the SIS model in continuous time.
At each time step, one randomly chosen active node deactives with
the probability $N_{i}/\left(N_{i}+\lambda N_{n}\right)$ where $N_{i}$
is the number of active nodes at time $t$, $N_{n}$ is the number
of edges emanating from them. With complementary probability $\lambda N_{n}/\left(N_{i}+\lambda N_{n}\right)$,
the active state is transmitted to one inactive neighbor of the randomly
selected node. Time is incremented by $\triangle t=1/\left(N_{i}+\lambda N_{n}\right)$.
This process is iterated after the system updating. 

\subsection{The Spectral Analysis for SIS Model\label{subsec:SIS-spectral}}

Here, we review the derivation of the criterion for the localization of steady active state on networks based on evaluation the inverse participation ratio
(IPR) of eigenvectors of the adjacency matrix. Denote the eigenvalues and eigenvectors of the adjacency matrix
$A_{xy}$ as $\Lambda_{i}$ and $f_{x}\left(\Lambda_{i}\right)$,
for which $\Lambda_{1}\geq\Lambda_{2}\geq\cdots\geq\Lambda_{N}$.
The probabilities $\rho_{x}$ at the steady state can be written as
a linear superposition of the $N$ orthogonal eigenvectors \citep{goltsev2012localization},
\begin{eqnarray}
\rho_{x} & = & \sum_{\Lambda}c\left(\Lambda\right)f_{x}\left(\Lambda\right).\label{eq:prob_decomposition}
\end{eqnarray}
If the largest eigenvalue $\Lambda_{1}$ is significantly larger than
all the other eigenvalues, i.e., there is a spectral gap in the spectrum,
then the QMF approximation predicts the critical point $\lambda_{c}$
to scale as $1/\Lambda_{1}$, and the steady state probability as
\begin{eqnarray}
\rho_{x} & \sim & c\left(\Lambda_{1}\right)f_{x}\left(\Lambda_{1}\right).\label{eq:prob_decomposition-approximation}
\end{eqnarray}
At the critical $\lambda_{c}$, the order parameter $\rho$, defined
as the average of active probability over all the nodes, can be expanded
as, 
\begin{eqnarray}
\rho & \sim & a_{1}\triangle+a_{2}\triangle^{2}+\ldots,\label{eq:prob_decomposition-expansion}
\end{eqnarray}
in which $\triangle=\lambda\Lambda_{1}-1\ll1$ with the coefficients
\begin{eqnarray}
a_{i} & = & \frac{\sum_{x=1}^{N}f_{x}\left(\Lambda_{i}\right)}{N\sum_{x=1}^{N}f_{x}^{3}\left(\Lambda_{i}\right)}.\label{eq:prob_decomposition-coefficient}
\end{eqnarray}

\begin{figure}
\centering
\begin{subfigure}{\columnwidth}
\centering
\includegraphics[height=6cm,width=\columnwidth]{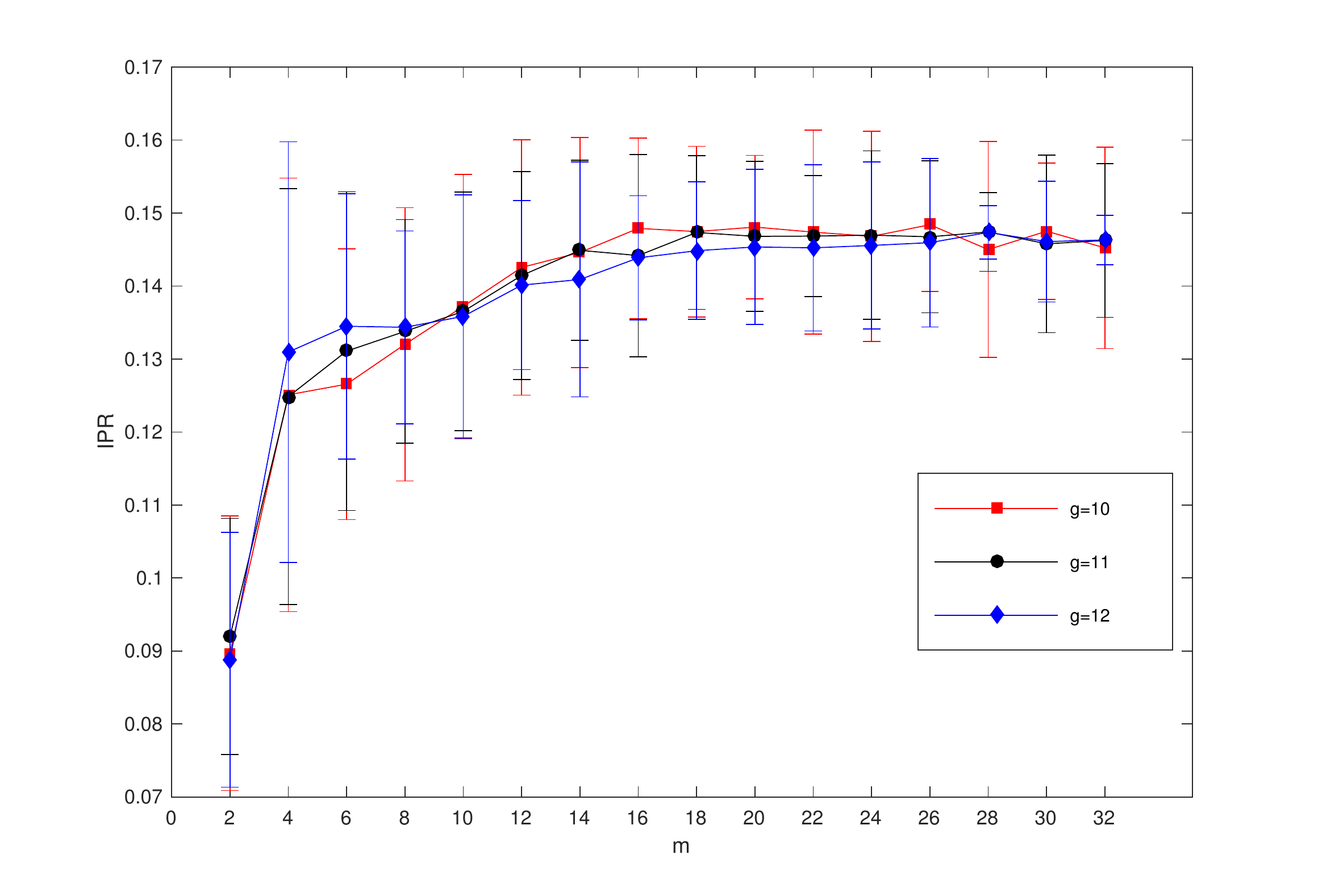}
\caption{}
\end{subfigure}
\begin{subfigure}{\columnwidth}
\centering
\includegraphics[height=6cm,width=\columnwidth]{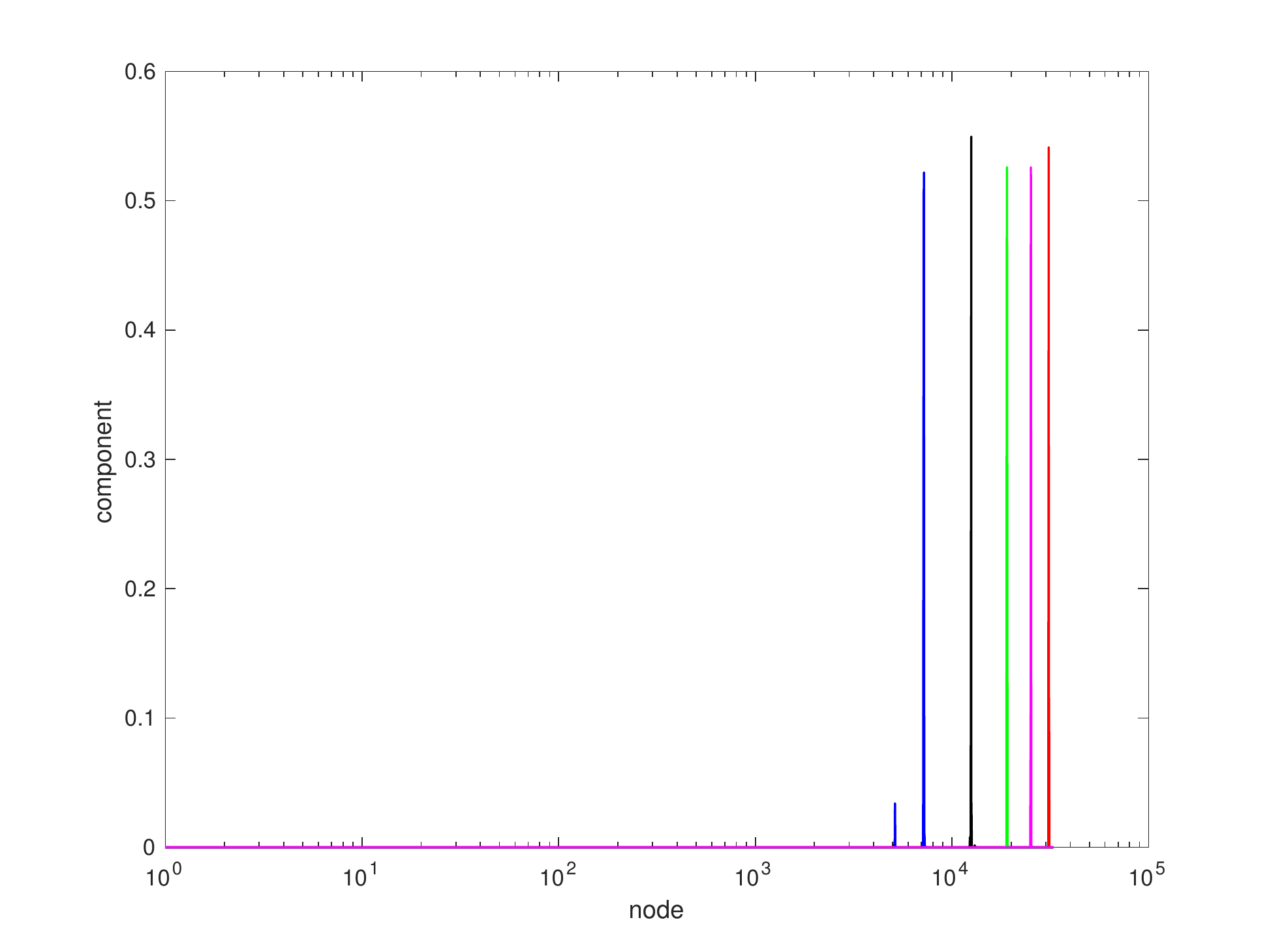}
\caption{}
\end{subfigure}

\captionsetup{justification=raggedright, singlelinecheck=false}
\caption{\label{fig:HN3-finite-IPR}(a) IPR vs $m$ for HMN1 of different generation $g$. The red squares are IPRs for different $m$ with $g=10$; the black circles are IPRs with $g=11$; the blue diamonds
are IPRs with $g=12$. Each data point averages IPRs over 100 independent realizations of HMN1. (b) the localized eigenvectors corresponding to five largest eigenvalues of the adjacency matrix of one graph realization of HMN1 for $g=11,m=16$.}
\end{figure}

With the dominant largest eigenvalue and the principle eigenvector,
the order parameter $\rho$ can be approximated with $\rho\sim a_{1}\triangle$.
In the limit $N\rightarrow\infty$, if the principle eigenvector $f_{x}\left(\Lambda_{1}\right)$
is localized, $a_{1}\sim O(1/N)$ and $\rho\sim O(1/N)$. Thus, the
active state is localized on the a few nodes of the network. In turn,
if the eigenvector $f_{x}\left(\Lambda_{1}\right)$ is delocalized
$(\sim\sqrt{N})$, $a_{1}\sim const$ and $\rho\sim const$. Then,
the active state extends over a finite fraction of nodes of the network.
As proposed in \citep{goltsev2012localization}, the localization
of the principle eigenvector is quantified by the inverse participation
ratio (IPR) of the principle eigenvector, 
\begin{eqnarray}
IPR\left(\Lambda\right) & = & \sum_{x=1}^{N}f_{x}^{4}\left(\Lambda\right).\label{eq:IPR}
\end{eqnarray}
A finite IPR corresponds to a localized principle eigenvector, while
a IPR approaching to zero corresponds to a delocalized principle eigenvector.
We apply the concept of IPR on all the networks we propose to examine
whether a localized principle eigenvector exists, which in the QMF approximation may suggest the 
the emergence of rare regions and the Griffiths phase \citep{moretti2013griffithsonHMNs,odor2015griffithsinHMNs}.

We analyze the principle eigenvector of the adjacency matrix of HMN1
for different generation $g$ with different size of basic cliques
$m$. As shown in Fig.(\ref{fig:HN3-finite-IPR}a),
the IPR increases with $m$ towards to a finite value. Each single value of IPR in Fig.(\ref{fig:HN3-finite-IPR}a) is derived by averaging over $100$ graph realizations of HMN1. 
Additionally, not only the largest eigenvalue, but actually a range of eigenvalues at the the higher edge in the spectrum have the localized eigenvectors, shown in Fig.(\ref{fig:HN3-finite-IPR}b).

\begin{figure}
\centering
\begin{subfigure}{0.9\columnwidth}
\includegraphics[height=5cm,width=\columnwidth]{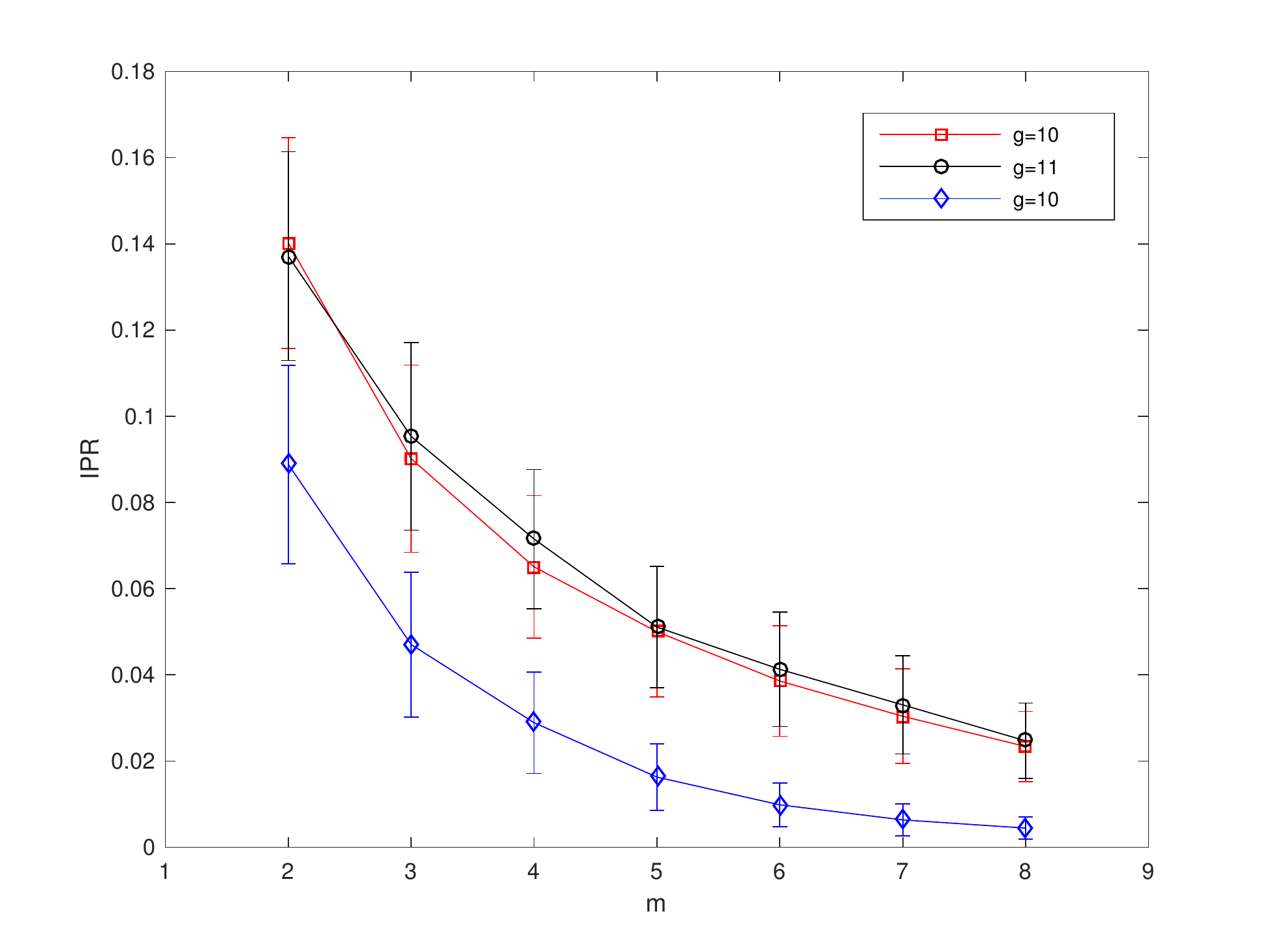}
\caption{}
\end{subfigure}
\begin{subfigure}{0.9\columnwidth}
 \includegraphics[height=5cm,width=\columnwidth]{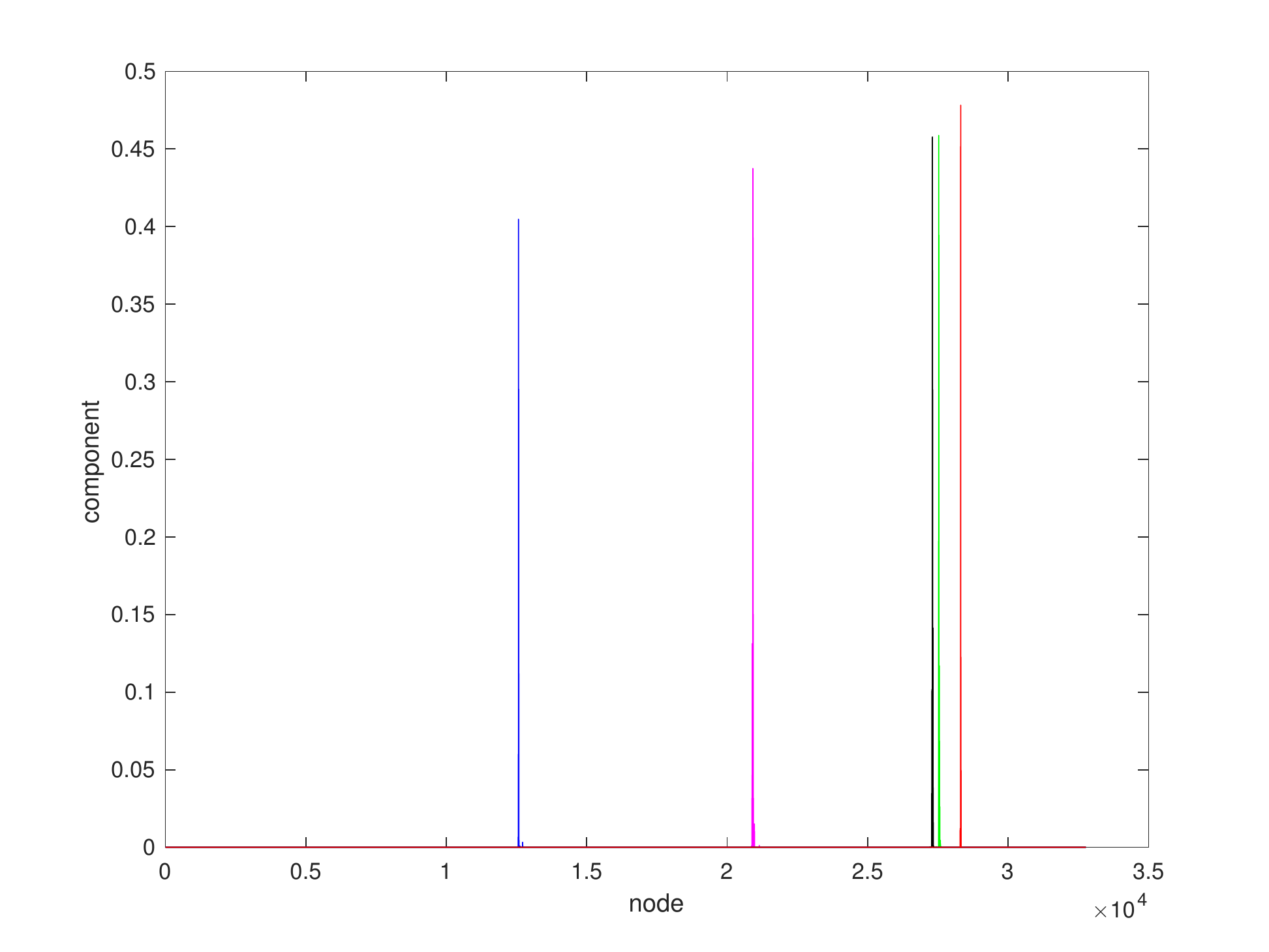}
 \caption{}
 \end{subfigure}
 \begin{subfigure}{0.9\columnwidth}  
 \includegraphics[height=5cm,width=\columnwidth]{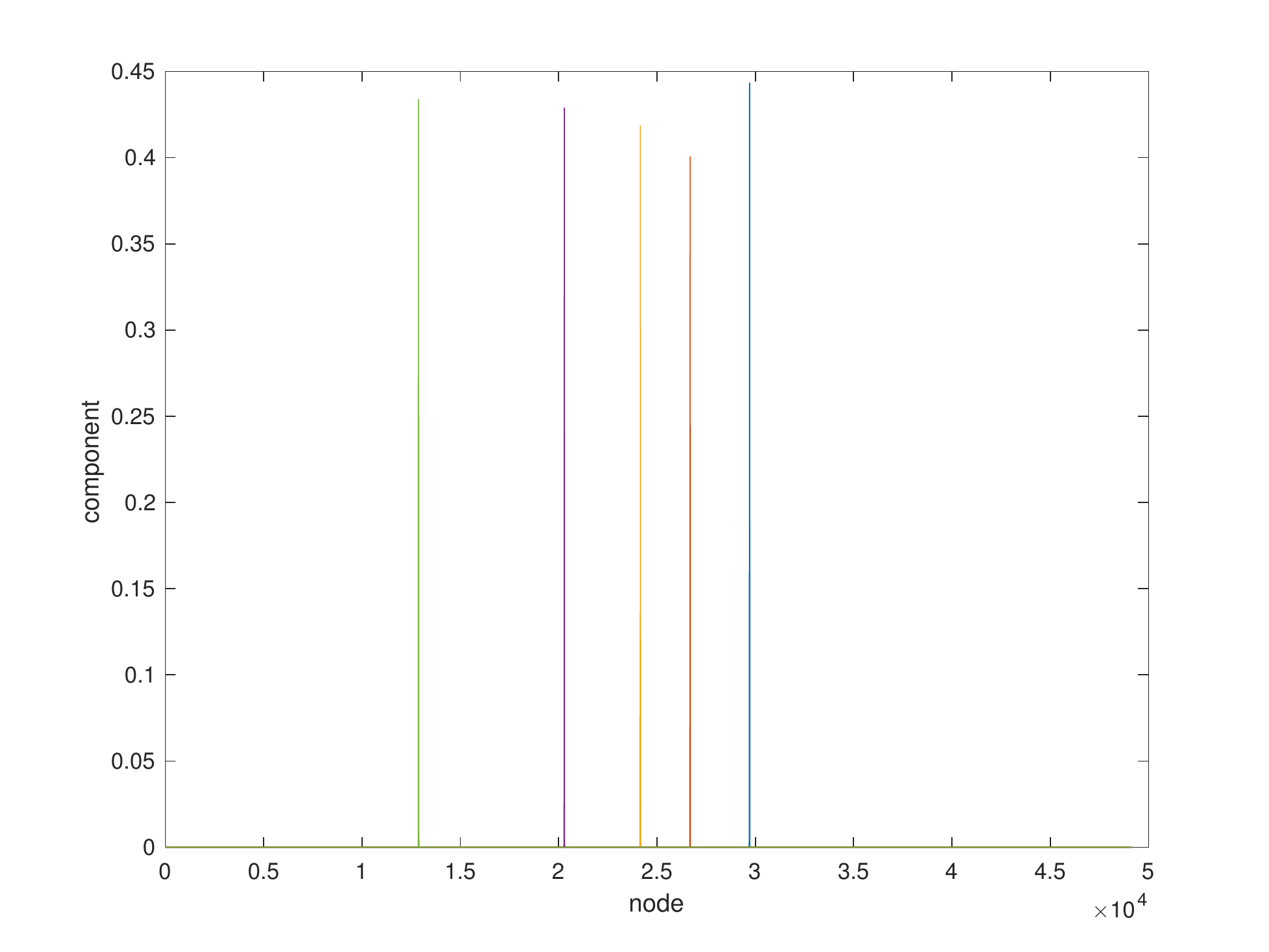}
\caption{}
\end{subfigure}   
\captionsetup{justification=raggedright, singlelinecheck=false}
\caption{\label{fig:HN3-HMN2-finite-IPR}(a)  IPR vs $m$  for HMN2 with different $g$. The red squares are values of the IPR with $g=10$; the black circles are IPRs for $g=11$.
The level-dependent inter-moduli probability is $p_{i}=4^{-\left(i+1\right)}$. Compared to them, the bottom line with blue diamonds is for $g=10$ with a level-dependent probability $p_{i}=4^{-i}$. Each data point averages IPRs over 100 independent realizations of HMN2. (b): localized
eigenvectors corresponding to five largest eigenvalues of the adjacency matrix of one graph realization of HMN2 with $g=14,m=2,p_{i}=4^{-\left(i+1\right)}$; (c): localized eigenvectors corresponding to five largest eigenvalues of the adjacency matrix of one graph realization of HMN2 with $g=14,m=3,p_{i}=4^{-\left(i+1\right)}$.}
\end{figure}

For HMN2, we work on simple level-dependent inter-moduli connectivity
probabilities, $p_{i}=4^{-\left(i+1\right)}$ and $p_{i}=4^{-i}$.
The backbone as well as the first hierarchy inter-moduli connectivity
probability is fixed at $1/4$, where the moduli are the basic cliques
described in Sec.\ref{sec:Graph-Structure}. The IPRs are shown in
Fig.(\ref{fig:HN3-HMN2-finite-IPR}a), from which we find the largest
IPR is from the the scheme that the single clique contains $2$ nodes,
and the probability is $p_{i}=4^{-\left(i+1\right)}$. In this case,
the network is statistically almost fragmented. Our numerical results
in Sec.\ref{sec:results} indeed show the emergence of the Griffiths
phase as a trivial consequence of the network disconectedness. To
obtain a connected network with a finite fractal dimension, we also
perform the simulation of the case in which $m=3$ and $p_{i}=4^{-\left(i+1\right)}$.
For HMN2, which is stochastically constructed, as the clique size
$m$ or level-dependent probability increases, the inter-moduli connections
become more and more dense and the IPR decreases, shown in Fig.(\ref{fig:HN3-HMN2-finite-IPR}a).
The regime over the parameter $m$ or the level-dependent $p_{i}$
for the possible emergence of the Griffiths phase is narrow. However,
the localized principle eigenvector exists for HMN2 with a finite
IPR. In Fig.(\ref{fig:HN3-HMN2-finite-IPR}b) and Fig.(\ref{fig:HN3-HMN2-finite-IPR}c),
we illustrate this result using two example graphs.

\section{Simulation Results for the SIS Model on HMN1 and HMN2\label{sec:results}}

\begin{figure}
\centering
\begin{subfigure}{0.45\textwidth}
\centering
\includegraphics[width=\textwidth]{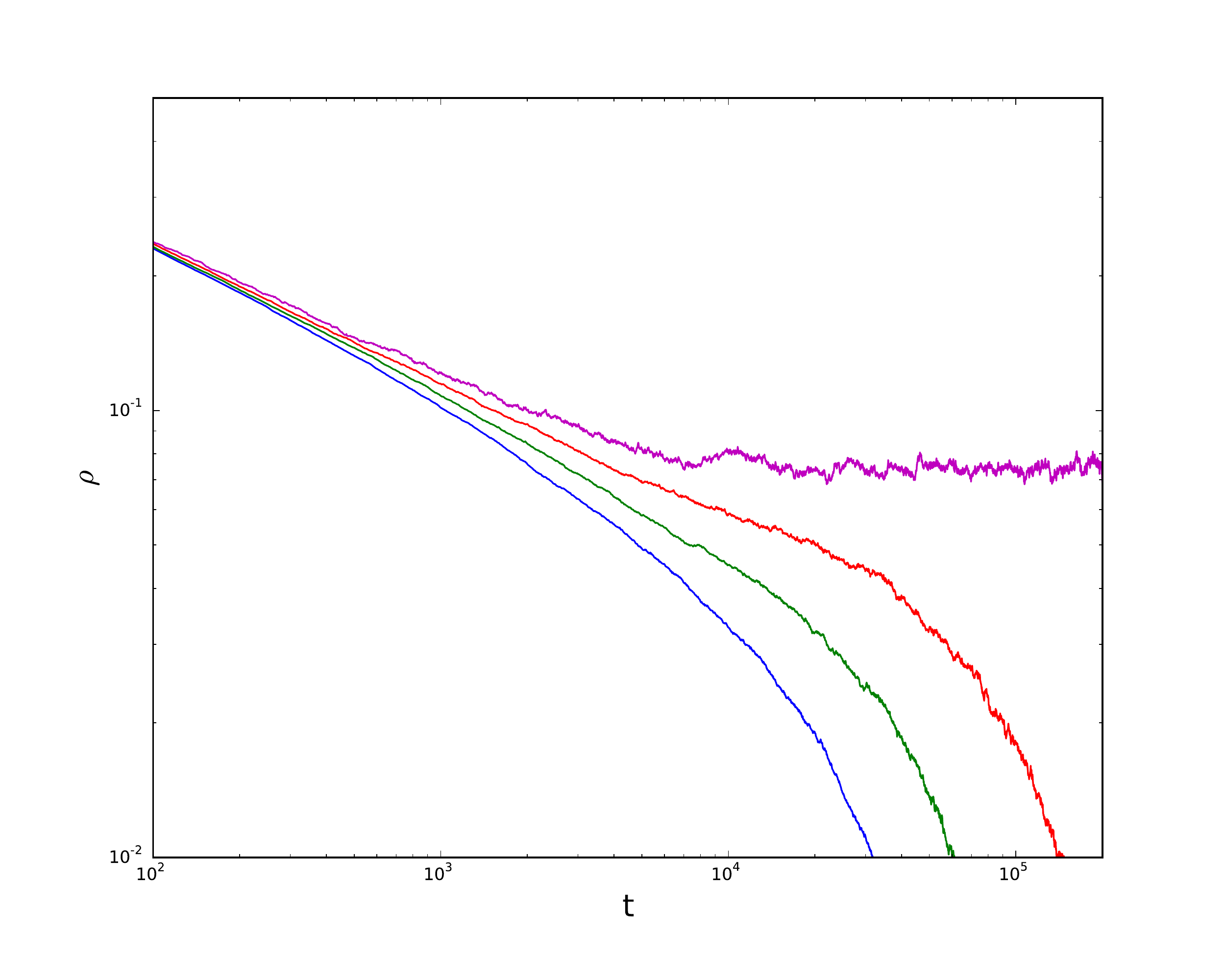}
\caption{}
\end{subfigure}
\begin{subfigure}{0.45\textwidth}
\centering
\includegraphics[width=\textwidth]{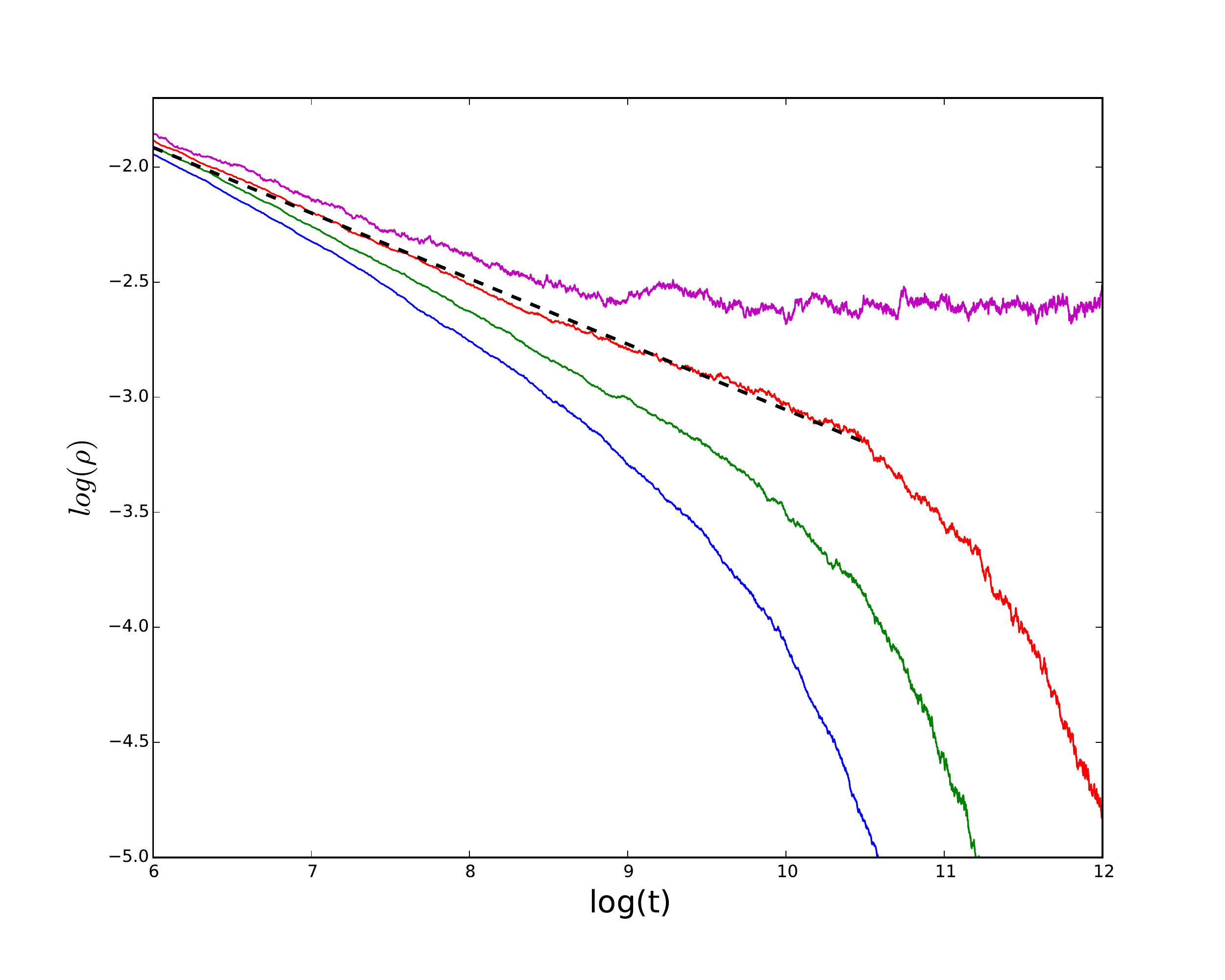}
\caption{}
\end{subfigure}

\captionsetup{justification=raggedright, singlelinecheck=false}
\caption{\label{fig:HMN1-11-8}(a): $\rho$ vs $t$ for HMN1 with $g=11,m=8$; blue line: $\lambda=0.1650$; green line: $\lambda=0.1651$; red line: $\lambda=0.1652$; magenta line:
$\lambda=0.1653$; (b): $\log\left(\rho\right)$ versus $\log\left(t\right)$ , the black dashed line is the fitted curve with $\rho\sim t^{-0.2849\ldots}$. The critical propagation rate is $\lambda_{c} \approx 0.1652$}
\end{figure}

In this section, we use the simulation method introduced in Sec.\ref{subsec:SIS-simulation} to run the SIS model on HMN1 and HMN2. The network is initialized as a fully-active graph. The system is updated each step until
$t_{max} \left(10^{6}\right)$ is reached or in case of activity extinction. Simulations for each propagation rate $\lambda$ are repeated for $1000 \sim 5000$ independent network realizations that are averaged over to obtain the order parameter
$\rho\left(t\right)$. We also derive the effective decay exponent by fitting critical power laws $\rho\left(t\right) \sim t^{-\alpha_{eff}}$ with (\citep{odor2013spectral,odor2015griffithsinHMNs})

\begin{eqnarray}
\alpha_{eff} & = -\frac{\ln\left[\rho\left(t\right)/\rho\left(t^{\prime}\right)\right]}{\ln\left(t/t^{\prime}\right)}
\end{eqnarray}

\begin{figure}
\centering
\begin{subfigure}{0.45\textwidth}
\centering
\includegraphics[width=\textwidth]{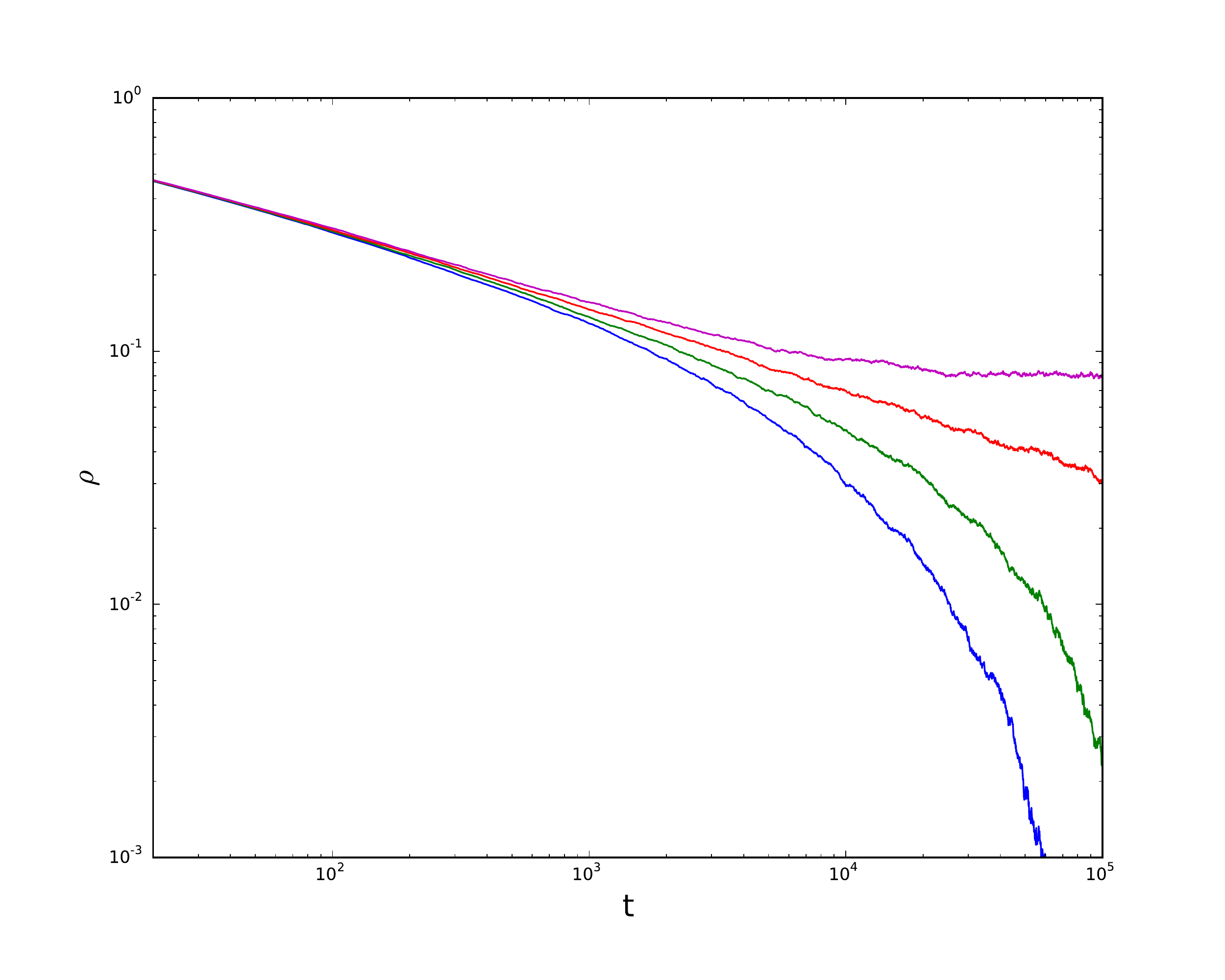}
\caption{}
\end{subfigure}
\begin{subfigure}{0.45\textwidth}
\centering
\includegraphics[width=\textwidth]{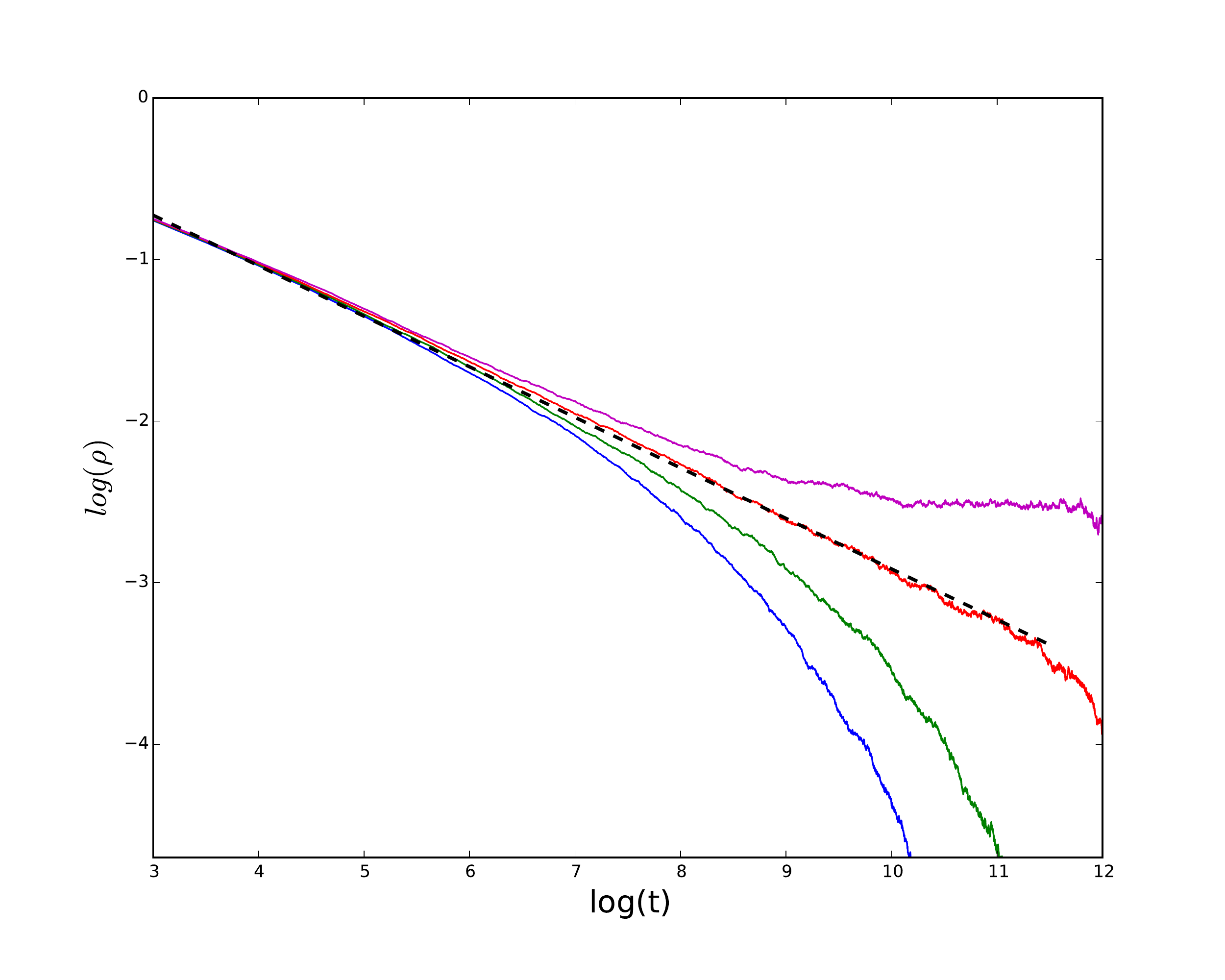}
\caption{}
\end{subfigure}

\captionsetup{justification=raggedright, singlelinecheck=false}
\caption{\label{fig:HMN1-11-16}(a):  $\rho$ vs $t$ for HMN1 with $g=11,m=16$; blue line: $\lambda=0.07475$; green line: $\lambda=0.0.07480$; red line: $\lambda=0.07485$; magenta
line: $\lambda=0.0749$; (b): $\log\left(\rho\right)$ versus $\log\left(t\right)$ , the black dashed line is the fitted curve with $\rho\sim t^{-0.3127\ldots}$. The critical propagation rate is $\lambda_{c}\approx0.07485$.}
\end{figure}

\begin{figure}
\centering
\begin{subfigure}{0.45\textwidth}
\centering
\includegraphics[width=\textwidth]{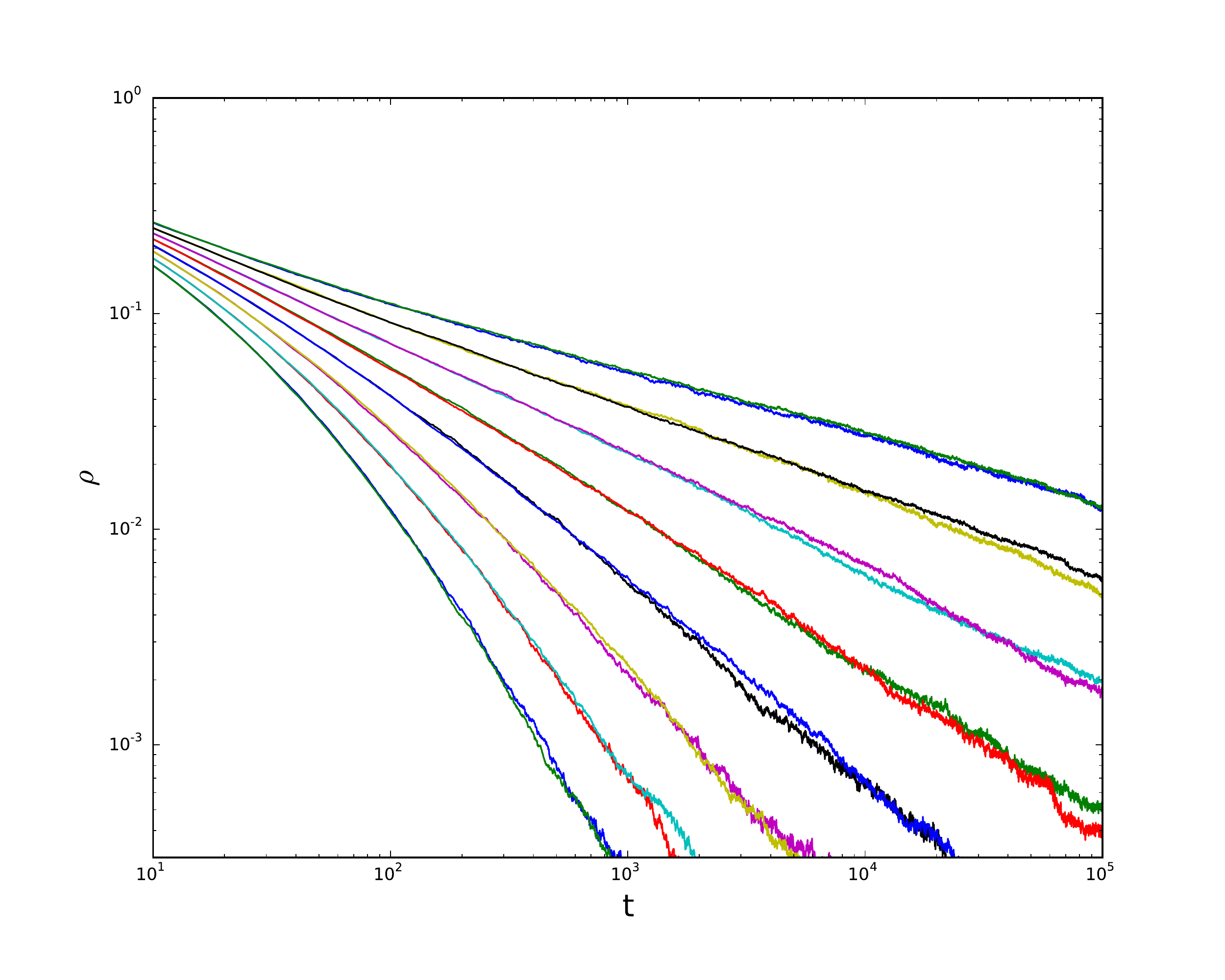}
\caption{}
\end{subfigure}
\begin{subfigure}{0.45\textwidth}
\centering
\includegraphics[width=\textwidth]{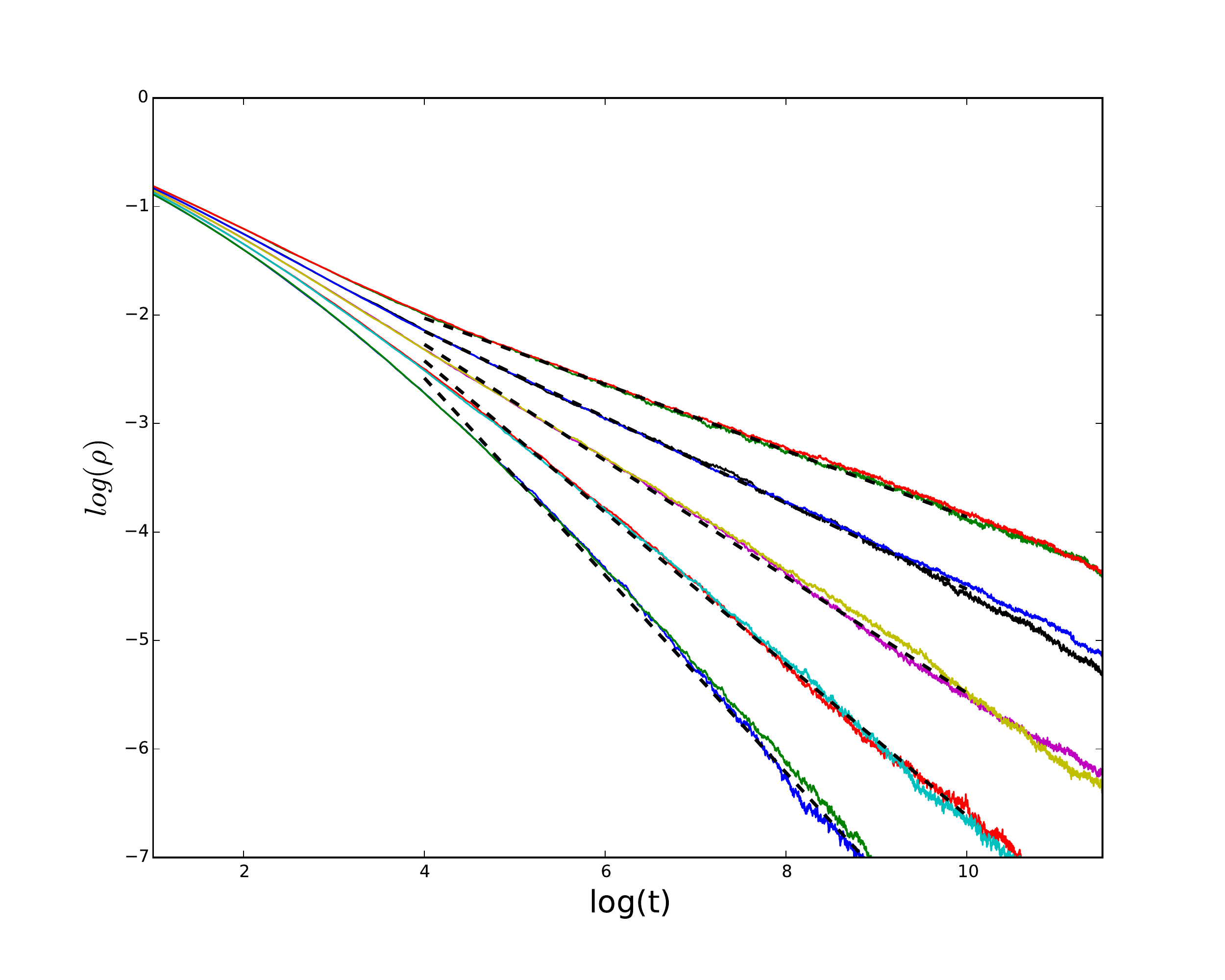}
\caption{}
\end{subfigure}

\captionsetup{justification=raggedright, singlelinecheck=false}
\caption{\label{fig:HMN2-13-14-2}(a): $\rho$ vs $t$ for HMN2 with $g=13,m=2$ and with $g=14,m=2$. Lines from bottom to top are for
$\lambda=0.46,0.47,0.48,0.49,0.50,0.51,0.52,0.53$. (b): $\log\left(\rho\right)$ versus $\log\left(t\right)$, the black dashed lines are the fitted curves with $\rho\sim t^{-0.9094\ldots}$, $\rho\sim t^{-0.6989\ldots}$, $\rho\sim t^{-0.5356\ldots}$, $\rho\sim t^{-0.3962\ldots}$
and $\rho\sim t^{-0.3054\ldots}$ from bottom to top for $\lambda=0.49,0.50,0.51,0.52,0.53$}
\end{figure}

\begin{figure}
\centering
\begin{subfigure}{0.45\textwidth}
\centering
\includegraphics[width=\textwidth]{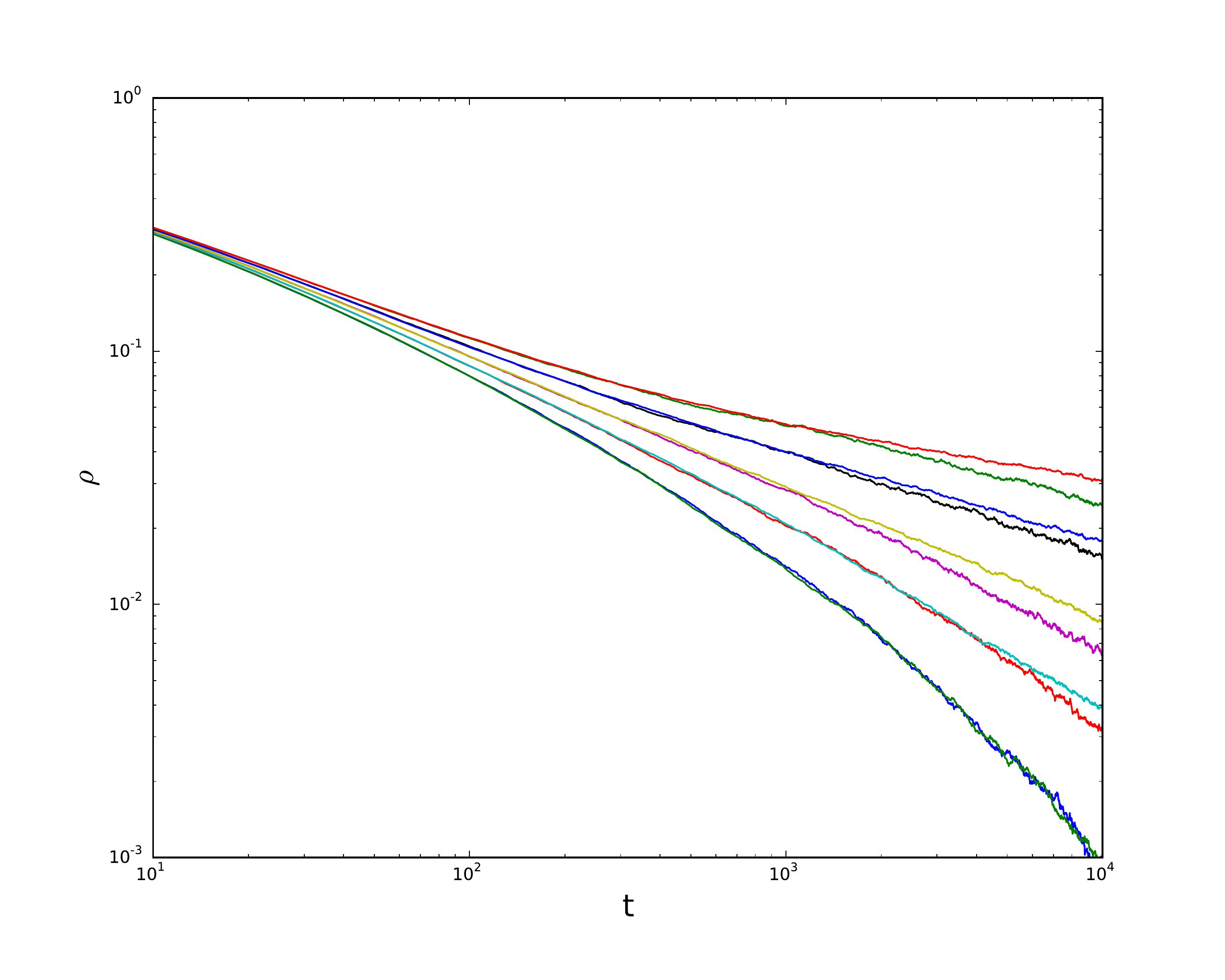}
\caption{}
\end{subfigure}
\begin{subfigure}{0.45\textwidth}
\centering
\includegraphics[width=\textwidth]{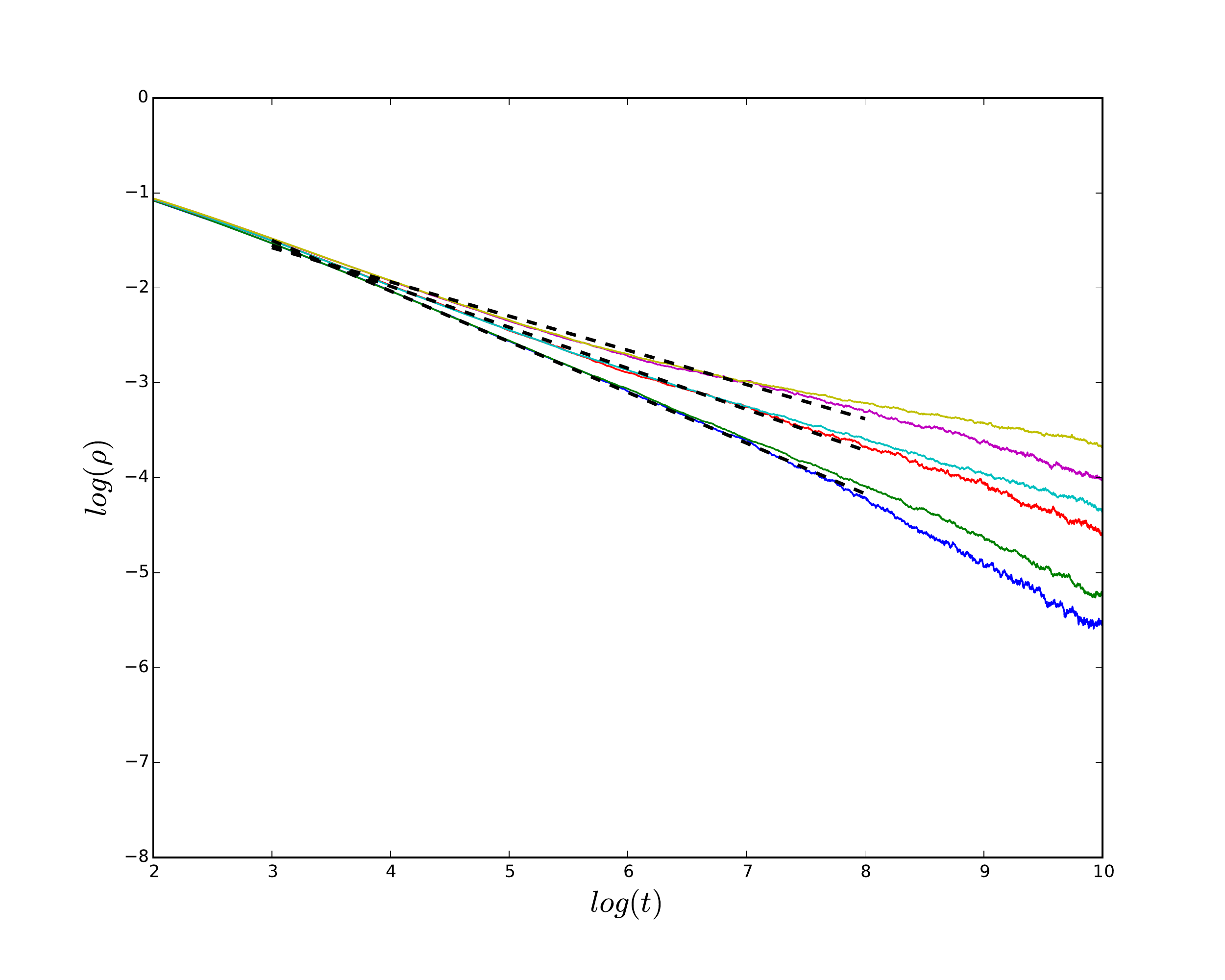}
\caption{}
\end{subfigure}

\captionsetup{justification=raggedright, singlelinecheck=false}
\caption{\label{fig:HMN2-12-13-3}(a): $\rho$ vs $t$ for HMN2 with $g=13,m=3$ and with $g=12,m=3$.  Lines from bottom to top are for propagation rates
$\lambda=0.258,0.259,0.260,0.261,0.262$. (b): $\log\left(\rho\right)$ versus $\log\left(t\right)$, the black dashed lines from bottom to top are the fitted curves with $\rho\sim t^{-0.5339\ldots}$, $\rho\sim t^{-0.4325\ldots}$, $\rho\sim t^{-0.3605\ldots}$ for $\lambda=0.260,0.261,0.262$}
\end{figure}

In Fig.(\ref{fig:HMN1-11-8}) and Fig.(\ref{fig:HMN1-11-16}), we present the simulation results of the SIS model on HMN1 with $g=11,m=8$ and $g=11,m=16$, and fit with the effective decay exponent at the critical point. The Griffiths phase is absent in in HMN1, and we find trivial phase transition at criticality. 

For HMN2 with $m=2$, the size-independent Griffiths phase emerges, shown in Fig.(\ref{fig:HMN2-13-14-2}). However, the Griffiths phase is a trivial consequence of the disconnectedness
of HMN2 when $p_{i}=4^{-\left(i+1\right)}$. We perform the simulation for HMN2 with $m=3,p_{i}=4^{-\left(i+1\right)}$ that is statistically almost certain to be connected. As the connections
are established stochastically, there is a chance that all the possible
inter-moduli edges fails to be connected. To avoid this case, we enforce
at least one inter-moduli connection to exist by repeating the construction
process in the simulation. The numerical results for a connected HMN2
is presented in Fig.(\ref{fig:HMN2-12-13-3}). We find a nearly size-independent
power laws in a stretched regime of $\lambda$. Comparing Fig.(\ref{fig:HMN2-13-14-2})
with Fig.(\ref{fig:HMN2-12-13-3}), we expect that, as $m$ increases
while keeping $p_{i}$ fixed and $\mathcal{\mathsf{\mathit{vice}}}$
$\mathit{versa}$, the regime in the parameter space of $\lambda$
for the Griffiths phase becomes narrow until it disappears when HMN2
becomes high-dimensional.

\section{Conclusion\label{sec:conclusion}}

In this work, we construct two classes of synthetic hierarchical modular
networks that possess a self-similar structure and small-world long
range connections, based on the hierarchical Hanoi networks \citep{boettcher2015real}.
We study the Griffiths phase by evolving the fundamental SIS model
on the HMNs we design. As an further exploration into a Griffiths
phase that is caused by the structural heterogeneity of networks,
we compare numerical results for two classes of networks. The
results suggest the essential role of the exponential distribution
of the inter-moduli connectivity probability or, equivalently, the
size of moduli on the emergence of a Griffiths phase. The first class of hierarchical networks, HMN1, are
not able to support the Griffiths phase, although they satisfy the structural
criteria, such as the finite fractal dimension,
the modular structure, the hierarchical heterogeneity. The second class of hierarchical networks, HMN2, are constructed
to possess a hierarchical pattern in the inter-moduli connectivity
probability and size of moduli, which therefore require a delicate tuning to maintain
a connected, finite dimensional network. This significant difference
in the design of hierarchical pattern results in the emergence of
the Griffiths phase.

As a complement  to the computational efforts, the spectral analysis proposed in the Quenched Mean Field approximation
suggests that a finite IPR of the principle eigenvector of the network
adjacency matrix can be considered as an indicator of the localization of activity that 
may result in the emergence of rare regions and the Griffiths
phase under certain circumstances. Although all the networks we consider prove
to have a finite IPR and localized eigenvectors corresponding to the
higher edge of the spectrum, only when the structural
disorder of inter-moduli connections is sufficient, the Griffiths phase appears.  As an extension to previous finite dimensional models that support the Griffiths phase with a localized principle eigenvector \citep{moretti2013griffithsonHMNs,odor2015griffithsinHMNs}, we find a class of finite dimensional networks with a localized principle eigenvector on which the Griffiths phase is absent. 
This raises questions on a more generalized theoretical analysis that applies to all the networks considered previously and currently.

\section{Acknowledgements}

I would like to thank Prof. Stefan Boettcher for helpful discussions. This work is supported by the NSF through grant DMR-1207431 is gratefully acknowledged.

\bibliographystyle{apsrev}
\bibliography{GP_SIS}

\end{document}